
%
%
\input harvmac
%
%
%
%
\ifx\answ\bigans
\else
\output={
  \almostshipout{\leftline{\vbox{\pagebody\makefootline}}}\advancepageno
}
\fi
%
%
%

%
%

%
%
\def\UCSD#1#2{\noindent#1\hfill #2%
\bigskip\supereject\global\hsize=\hsbody%
\footline={\hss\tenrm\folio\hss}}
%
%
\def\abstract#1{\centerline{\bf Abstract}\nobreak\medskip\nobreak\par #1}
%
%
%
%
\edef\tfontsize{ scaled\magstep3}
 \tfontsize  \tfontsize
 \tfontsize \font\titlei=cmmi10 \tfontsize
\font\titleis=cmmi7 \tfontsize \font\titleiss=cmmi5 \tfontsize
\font\titlesy=cmsy10 \tfontsize \font\titlesys=cmsy7 \tfontsize
\font\titlesyss=cmsy5 \tfontsize  \tfontsize
\skewchar\titlei='177 \skewchar\titleis='177 \skewchar\titleiss='177
\skewchar\titlesy='60 \skewchar\titlesys='60 \skewchar\titlesyss='60
%
%
%
%
%
\def\inv{^{\raise.15ex\hbox{${\scriptscriptstyle -}$}\kern-.05em 1}}
\def\lbar{{\lower.35ex\hbox{$\mathchar'26$}\mkern-10mu\lambda}} 

%
%
%
%
\def\dsl{\,\raise.15ex\hbox{/}\mkern-13.5mu D} 
\def\delsl{\raise.15ex\hbox{/}\kern-.57em\partial}
\def\Ksl{\hbox{/\kern-.6000em\rm K}}
\def\Asl{\hbox{/\kern-.6500em \rm A}}
\def\Dsl{\hbox{/\kern-.6000em\rm D}} 
\def\Qsl{\hbox{/\kern-.6000em\rm Q}}
\def\gradsl{\hbox{/\kern-.6500em$\nabla$}}
%
%
\def\lspace{\ifx\answ\bigans{}\else\qquad\fi}
\def\lbspace{\ifx\answ\bigans{}\else\hskip-.2in\fi} 
%
%
\def\boxeqn#1{\vcenter{\vbox{\hrule\hbox{\vrule\kern3pt\vbox{\kern3pt
        \hbox{${\displaystyle #1}$}\kern3pt}\kern3pt\vrule}\hrule}}}
%
%
\def\mbox#1#2{\vcenter{\hrule \hbox{\vrule height#2in
\kern#1in \vrule} \hrule}}
%
%
%
%

  \def\CO{{\cal O}}

%
%
%
%
%

%

\def\bar#1{\overline{#1}}

\def\bra#1{\left\langle #1\right|}
\def\ket#1{\left| #1\right\rangle}

\def\darr#1{\raise1.5ex\hbox{$\leftrightarrow$}\mkern-16.5mu #1}

%
%
\def\frac#1#2{{\textstyle{#1\over #2}}} 
%
%
%
%

%
%
%
%

%
%
\def\ltap{\ \raise.3ex\hbox{$<$\kern-.75em\lower1ex\hbox{$\sim$}}\ }
\def\gtap{\ \raise.3ex\hbox{$>$\kern-.75em\lower1ex\hbox{$\sim$}}\ }
\def\gl{\ \raise.5ex\hbox{$>$}\kern-.8em\lower.5ex\hbox{$<$}\ }
\def\roughly#1{\raise.3ex\hbox{$#1$\kern-.75em\lower1ex\hbox{$\sim$}}}
%
%

%

%
\def\np#1#2#3{Nucl. Phys. B{#1} (#2) #3}
\def\pl#1#2#3{Phys. Lett. {#1}B (#2) #3}
\def\prl#1#2#3{Phys. Rev. Lett. {#1} (#2) #3}

\relax

\noblackbox
\def\clebsch#1#2#3#4#5#6{\left(\left.
\matrix{{#1}&{#2}\cr{#4}&{#5}\cr}\right|\matrix{{#3}\cr{#6}}\right)}

\def\nc{N_c}
\def\ln{large-$\nc$}

\centerline{{\titlefont{$1/\nc$ Corrections to the}}}
\centerline{{\titlefont{Baryon Axial Currents in QCD}}}
\bigskip
\centerline{Roger Dashen and Aneesh V.~Manohar}
\smallskip
\centerline{{\sl Department of Physics, University of California at San
Diego, La Jolla, CA 92093}}
\bigskip
\vfill
\abstract{We prove that the $1/\nc$ corrections to the baryon axial current
matrix elements are proportional to their lowest order values. This implies
that
the first correction to axial current coupling constant ratios vanishes, and
that the $SU(2N_f)$ spin-flavor symmetry relations are only violated at second
order in $1/\nc$. }
\vfill
\UCSD{\vbox{\hbox{UCSD/PTH 93-18}\hbox{hep-ph/9307242}}}{July 1993}

In the \ln\ limit of QCD~\ref\thooft{G. 't Hooft, \np{72}{1974}{461}}, the
lowest lying baryon states form an infinite tower of states with
$I=J=1/2, 3/2, \ldots$. The matrix element of the axial current between
baryon states is of order $\nc$, and can be written as
\eqn\axialmatrix{
\bra{N} \bar\psi\gamma^i\gamma_5 \tau^a\psi \ket{N} = g_0 \nc \bra{N} X^{ia}_0
\ket{N},
}
where $X^{ia}_0$ and $g_0$ are of
order one. In a previous paper~\ref\dm{R. Dashen and A.V. Manohar,
UCSD/PTH 93-16}, we showed how unitarity in pion-baryon scattering and
consistency of chiral perturbation theory implied that the infinite tower of
baryon states must be degenerate, and that the axial couplings must satisfy
\eqn\cons{
\left[X^{ia}_0,X^{jb}_0\right]=0,
}
a result also obtained previously by Gervais and Sakita~\ref\gs{J.-L.
Gervais and B.~Sakita, \prl{52}{1984}{87}, Phys.~Rev.~D30 (1984) 1795}. This
relation implies that the baryon states form a representation of a contracted
$SU(4)$ spin-flavor algebra. The matrix elements of $X^{ia}_0$ can be written
in
terms of reduced matrix elements $X_0(J,J^\prime)$ as \eqn\reduced{
\bra{J^\prime,m^\prime,\alpha^\prime} X^{ia} \ket{J,m,\alpha}
= X_0(J,J^\prime) \sqrt{{2J+1\over 2J^\prime+1}} \clebsch J 1 {J^\prime} m i
{m^\prime} \clebsch J 1 {J^\prime} \alpha a {\alpha^\prime},
}
where the normalization constant has been chosen so that
$X_0(J,J^\prime)=X_0(J^\prime,J)$. The constraint eq.~\cons\ implies
that all the reduced matrix elements are equal. The choice
$X_0(J,J^\prime)=1$ for the normalization  fixes the normalization of $g_0$ in
eq.~\axialmatrix. In this paper, we will compute the $1/\nc$ correction to
the axial current matrix elements. We will show that the order one
contributions to the matrix element of the axial current between baryon states
is proportional to the leading order $\nc$ contribution. Thus the pion-baryon
coupling constant ratios in the baryon tower only get a correction at order
$1/\nc^2$. This result is important because it simplifies the
computation of the $1/\nc$ corrections of the weak and electromagnetic
properties of baryons~\ref\djm{R. Dashen, E. Jenkins, and A.V. Manohar,
UCSD/PTH 93-21}.

In studying the $1/\nc$ corrections, it is important to always
work in the limit that the isospin $I$ and spin $J$ of the
baryon are held fixed as $\nc\rightarrow\infty$. For $\nc=3$, the $I=J$
tower of baryons is not infinite, but stops at the $\Delta$ resonance
with $I=J=3/2$. The finite height of the tower is
automatically included as $1/\nc$ operators in the effective
hamiltonian. For small $I$ and $J$, the $1/\nc$ corrections will be
under control. As $I$ and $J$ get larger, the corrections increase, and
eventually the $1/\nc$ expansion breaks down. The top of the baryon tower for
finite $\nc$ is the point at which the $1/\nc$ expansion breaks down.
Clearly, the larger the value of $\nc$, the larger the values of $I$ and $J$
before the perturbation expansion breaks down.

Consider the scattering process
$\pi^a+N\rightarrow \pi^b+\pi^c+N$ at low energies. The nucleon pole graphs
that
contribute in the \ln\ limit are shown in \fig\pppn{Diagrams contributing to
$\pi N\rightarrow \pi\pi N$. Graphs (c) and (d) are suppressed by $1/\nc^2$
relative to the leading term.}. Graphs with multiple pions at the same vertex,
such as those in figs.~1(c) and (d) are suppressed by $1/\nc^2$ relative to the
leading terms. Define the matrix element of the axial current to order $1/\nc$
by
\eqn\amatrix{ \bra{N} \bar\psi\gamma^i\gamma_5 \tau^a\psi \ket{N} = g_0 \nc
\bra{N} X^{ia} \ket{N}, }
where $g_0$ is a constant independent of $\nc$. $X^{ia}$ then can be
expanded in a series in $1/\nc$,
\eqn\gexp{
X^{ia}= X^{ia}_0 +{1\over\nc} X^{ia}_1+\ldots
}
The amplitude for pion-nucleon scattering from the
diagrams in \pppn\ is proportional to
$$
\nc^{3/2} \left[X^{ia},\left[X^{jb},X^{kc}\right]\right]
$$
and violates unitarity unless the double commutator vanishes at least as
fast as $\nc^{-3/2}$, so that the amplitude is at most of order one. (In fact,
one expects that the double commutator is of order $1/\nc^2$ since the
corrections should only involve integer powers of $1/\nc$. This result also
follows from the \ln\ counting rules which imply that each additional pion has
a factor of $1/\sqrt{\nc}$ in the amplitude.) Substituting eq.~\gexp\ into the
constraint gives  \eqn\acons{
\left[X^{ia}_0,\left[X_1^{jb},X^{kc}_0\right]\right] +
\left[X^{ia}_0,\left[X^{jb}_0,X_1^{kc}\right]\right] = 0, }
using $\left[X^{ia}_0,X^{jb}_0\right]=0$. The only solution to the
consistency equation~\acons\ is that $X_1^{ia}$ is proportional to
$X_0^{ia}$. This can be verified by an explicit computation writing $X_1^{ia}$
in terms of reduced matrix elements, or by using group theoretic methods
discussed in ref.~\djm. Thus we find that
\eqn\gans{
X^{ia} = \left(1 + {c\over \nc}\right) X_0^{ia} + \CO\left({1\over
\nc^2}\right)
}
where $c$ is an unknown constant. The first correction to $X^{ia}$ is
proportional to the lowest order value $X_0^{ia}$,  so the $1/\nc$ correction
to
the axial coupling constant ratios vanishes. The overall normalization factor
$(1+c/\nc)$ can be reabsorbed into a redefinition of the unknown axial coupling
$g_0$ by the rescaling $g_0 \rightarrow g = g_0 (1+c/\nc)$, $X^{ia} \rightarrow
X_0^{ia}$, so there are
no new parameters at order $1/\nc$ in the axial current sector.

Eq.~\gans\  implies that
the double commutator $\left[X^{ia},\left[X^{jb},X^{kc}\right]\right]$ is of
order $1/\nc^2$, since the $1/\nc$ terms in the expansion of the double
commutator vanish. The one-loop renormalization of the pion-baryon couplings
from the graphs in \fig\aren{Diagrams contributing to the one-loop corrections
of the pion-baryon couplings.}\ is proportional to $\nc
\left[X^{ia},\left[X^{ia},X^{jb}\right]\right]$. The cancellation in the double
commutator implies that the one-loop radiative correction is of order $1/\nc$,
rather than $\nc$. Thus pion-loop
effects are supressed by $1/\nc$ in both the meson and the baryon sector.

The vanishing of the $1/\nc$ corrections to the axial couplings is
similar to the Ademollo-Gatto theorem~\ref\agatto{M. Ademollo and R. Gatto,
\prl{13}{1964}{264}}, and to the vanishing of $1/m_Q$ corrections in the heavy
quark theory at zero recoil~\ref\luke{M.E.
Luke, \pl{252}{1990}447\semi C.G. Boyd and D.E. Brahm, \pl{257}{1991}{393}}.
There are two sources of $1/\nc$ corrections, $1/\nc$
corrections to the states, and $1/\nc$ corrections to the axial current
operator. The $\pi-N$ scattering argument shows that the sum of these
$1/\nc$ corrections is proportional to the lowest order axial current matrix
element. This differs from the heavy quark theory, in which the $1/m_Q$
corrections vanish at zero recoil. In the heavy quark theory, the vector
current is a symmetry of the full theory for any $1/m_Q$, and normalizes the
form factors at zero recoil. We only have a contracted
$SU(4)$ symmetry of the effective theory in the limit $\nc\rightarrow\infty$,
with $\left[X^{ia}_0,X_0^{jb}\right]=0$. Thus the normalization of
$X_0^{ia}$ cannot be determined from the commutation relations. What we have
shown by studying $\pi-N$ scattering is that the $1/\nc$ corrections to the
commutation relations of the contracted $SU(4)$ algebra vanish. This implies
that the only possible $1/\nc$ correction is a rescaling of  $X^{ia}_0$.

The non-relativistic quark model provides a representation of the $SU(4)$
algebra for finite $\nc$, and it is easy to verify that
$\left[X^{ia},X^{jb}\right]$ is of order $1/\nc^2$ using eq.~(11) and (12) of
ref.~\dm. This implies that $1/\nc$ corections to the axial currents in the
non-relativistic quark model are proportional to their lowest order
values, a result that can also be obtained by  explicit computation.
There is no reason, however, why the quark model value for the $1/\nc$
correction (the value $c$ in eq.~\gans) should be the same as the QCD value.
The
quark model predictions for the ratios of pion-baryon couplings differs from
those of QCD only at second order in the $1/\nc^2$. This explains why the quark
model predictions for the coupling constant ratios agree so well with
experiment. The Skyrme model also gives a $1/\nc$ expansion of the axial
currents, in which the first correction to the coupling constant ratios is of
order $1/\nc^2$. There is no reason why the Skyrme model value for the
$1/\nc$ correction should agree with QCD. The \ln\ limit of QCD predicts that
pion-baryon couplings should respect $SU(4)$ spin-flavor symetry. We have seen
that the first correction to this result vanishes, which explains why the
experimental pion-baryon coupling ratios are close to their $SU(4)$-symmetric
values.

We would like to thank E.~Jenkins for helpful discussions. One of us (A.M.)
would like to thank the Aspen Center for Physics for hospitality while this
work was completed. This work was supported in part by DOE grant
DOE-FG03-90ER40546 and by PYI award PHY-8958081.

\listrefs
\listfigs
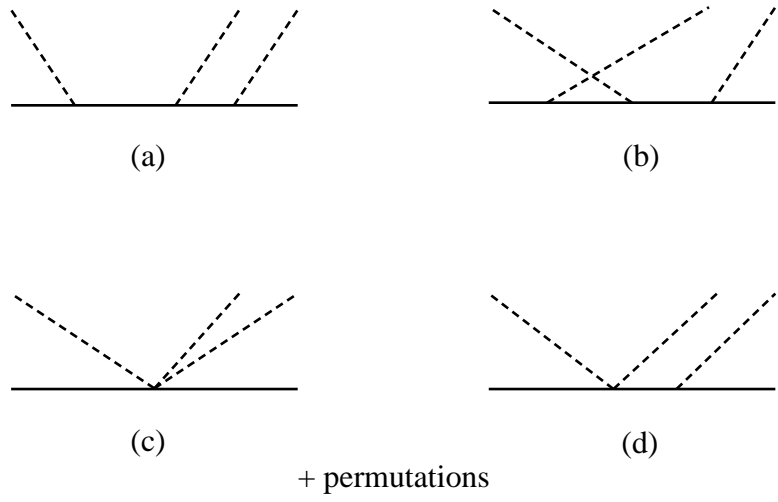
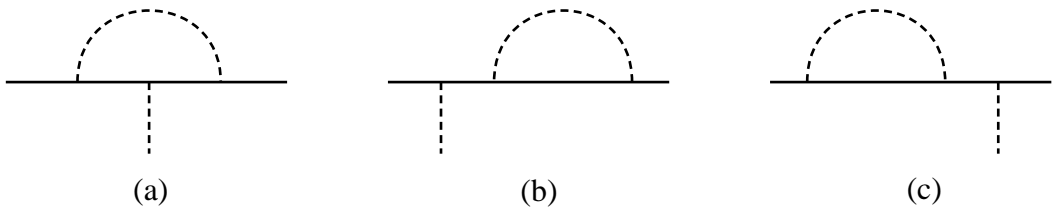
\bye